\documentclass{jdmdh}
\usepackage{array}
\usepackage{pgfplots}

\title{How the Taiwanese Do China Studies: Applications of Text Mining}
\author[1]{Hsuan-Lei Shao}
\author[2]{Sieh-Chuen Huang}
\author[3]{Yun-Cheng Tsai}
\affil[1]{Department of East Asian Studies, National Taiwan Normal University} 
\affil[2]{College of Law, National Taiwan University}
\affil[3]{Center for General Education, National Taiwan University}

\corrauthor{Yun-Cheng Tsai}{pecutsai@ntu.edu.tw}

\begin{document}

\maketitle

\abstract{With the rapid evolution of cross-strait situation, ``Mainland China'' as a subject of social science study has evoked the voice of ``Rethinking China Study'' among intelligentsia recently.
This essay tried to apply an automatic content analysis tool (CATAR) to the journal ``Mainland China Studies'' (1998-2015) in order to observe the research trends based on the clustering of text from the title and abstract of each paper in the journal.
The results showed that the 473 articles published by the journal were clustered into seven salient topics.
From the publication number of each topic over time (including ``volume of publications'', ``percentage of publications''),
there are two major topics of this journal while other topics varied over time widely. The contribution of this study includes:
\begin{enumerate}
\item We could group each ``independent'' study into a meaningful topic, as a small scale experiment verified that this topic clustering is feasible.
\item This essay reveals the salient research topics and their trends for the Taiwan journal ``Mainland China Studies''.
\item Various topical keywords were identified, providing easy access to the past study.
\item The yearly trends of the identified topics could be viewed as signature of future research directions.
\end{enumerate}}


\section{Introduction}

\strut
\vspace{-4ex}

Nowadays, digital method is penetrating into other field broadly. Even in the fields that we think those are not ``digital things'', such as philosophy, literature and so on. It is called digital humanities (DH). With regard to the theory of digital humanities, there has been considerable discussion nowadays. Besides the applications on each specific fields, DH researchers have gradually been willing to accept the ``text structure can be regarded as a numerical data structure'' (see~\cite{grimmer2013text}).

Then use data science methods to calculate documents. It is usually acceptable that ``computers can reduce the labour of researchers and
open up new insights for them'' at least. Therefore, method of text mining is to assist researchers rather than replace them in theoretically. Although there are different scale on how far can humanities phenomenon DH by data science method, Grimmer's viewpoint is still the maximum consensus. Recent researcher has included extending the use of databases and visual plug-ins to study classical Chinese books (see~\cite{doi101093llcfqx024}) or the social networks of a celebrity (see~\cite{finegold2016six}) and to capture article conversations onto novels (see~\cite{muzny2017dialogism}). They are all admit the principle of ``text as data'' basically.

Much of our research aim to understand which ideas Taiwanese researchers have been eager to know ``who studies China on what, when, and how'' (as the classical political question: ``Who Gets What When and How'', (see~\cite{lasswell1936gets}). To that end, we use text mining technology
to analyze published research articles according to their co-occurring words. This will help those who are interested in to have a whole picture in their field. The research described below is about the field of ``China Studies in Taiwan'' (CSIT). This field is a natural subject for text mining not only the subject is familiar to the authors that we could have sufficient domain knowledge to judge the machine learning models, but also it is undeniably the most important and dynamic influence on Taiwan for now and the near future. As the relationship between mainland China and Taiwan has gradually shifted, the academic circle in Taiwanese has had a growing tendency to put more emphasis on China Studies as it reflects on recent years. About the preliminary research, we have conducted is described below, followed by directions for further research. The contributions of this line of research include the following:
\begin{enumerate}
\item These studies reveal the salient research topics and trends in ``Mainland China Studies'', an academic journal edited by National Chengchi University. It is a typical journal in the China Studies field.
\item Various topical keywords were identified, providing future researchers easy access to the most important information published in Mainland China Studies, without having to read or even skim every individual article.
\item The yearly trends of the identified topics can provide valuable insights to guide future research directions.
\item More broadly, categorizing and grouping independent studies according to meaningful topics can provide an objective birds-eye view of the entire field of CSIT. The main propose of this paper is how data-science method can work/help in a specific academic area. In this manuscript, we also do it in ``short text in Chinese''. Our target in
this stage is to connect ``DH (China-studies)'' and ``DM (we use a digital tool)''. This is the spirit of ``CS+X'', which encourage we use some non-traditional method to our own domain knowledge field.
\end{enumerate}

\section{Preliminaries}
Mainland China is a popular subject of social science research. It has gradually gained significance in Taiwanese academic circle due to its special historical, political and economic relations throughout the world. The political status and social environment of the mainland have
changed tremendously since it began its "Reform-and-Opening-Up" policies in the late 1970s, making it a trendy topic around the globe. When it comes to the field of CSIT, Taiwanese scholars offer unique strengths. For one thing, it can be easier for Taiwanese scholars with a solid foundation in Western social science theories to perceive the development and the transformation of China's politics and society from a more objective angle than for many scholars from mainland China who do not rely on the same foundation.
In addition, scholars from Taiwan have innate linguistic advantages compared with most Western scholars because the official language in both mainland China and Taiwan is Chinese. As a result, Taiwanese scholars have played an important role in this field while the relationship
between Mainland China has got more important weight worldwide in recent years. In the 1990s, social scientists of CSIT began paying more attention to the integration of various methods and approaches within academia. During and since this decade, there has been a radical shift in the approach to CSIT. Such development includes, among other things, historical research and paradigm research. The first approach, historical research, marks the characteristics of each period of CSIT. Historical Speaking, CSIT could divide into three periods: Firstly, the banditry studies (before 1970's) have taken Chinese Communist Party (CCP) as an enemy, and focusing on military and war. Secondly, the CCP-centered studies (1970's to 1990's) have focused on factions' struggle in CCP, as Kremlinology in the US. Thirdly, we call it the Mainland China studies period (after 1990's) which is using modern social scientifical method to study.

Nowadays, CSIT has evolved from a focus on combating Communism to interest in business, social, and cultural issues; that is, it has largely shifted from hard politics to soft politics. The social scientifical research conducted through the lens of various disciplines (see~\cite{BaoWu, ChenGen}). In consultation with scholars from different disciplines, they examine China through a variety of theoretical models, including through the theory of integration, the divided nation model, the negotiation model, economic and trade policy, psychological analysis, strategic triangular theory, international system theory, safety code theory and so on. One of the purposes of these books is to provide examples of different research methods in CSIT. But a drawback of paradigm research is that it often presents a static picture, rather than revealing changes in China and in CSIT over time. Objectively, our research falls somewhere between and touches upon both of these approaches. In recent, there are many social science research on text mining or machine learning. For example, Researchers used structural topic model (STM) to help human-hand encoding more effectively (see~\cite{roberts2014structural}). Another paper showed the classifier machine in the political context (see~\cite{burscher2015using}).

On the hand, algorithms become more complex and more accurate.
There are researchers who used support vector machine (SVM) and Support Distribution Machine (SDM), and Latent Dirichlet Allocation (LDA) (see~\cite{boecking2015event}).
On the other hand, data scare is growing up as well, Wilkerson and Casas did a good exercise in counting big data (see~\cite{wilkerson2017large}).

\section{Methodologies}
In theory, most of the DH research papers have not claimed that they have ``a theoretical'' breakthrough or are trying to establish a ``new theory of DH''. The above essays are representative of the three trends in the development of DH today: Firstly, to use of data and math-tools to prove existing theories. Such as who visualize six degrees of social network theory (see~\cite{muzny2017dialogism}). Secondly, to give proof of empirical data academic arguments that can only remain in the concepts in the past, such as who proves the relationship between the ``Confucianism, Daoism, Legalism'' among Chinese classics (see~\cite{doi101093llcfqx024}).

This question could only be explained by the researchers' ``intuition'' before. Third, the invention of new technologies, such as who can achieve higher efficiency and draw on ``dialogue'', which focuses on technology (see~\cite{muzny2017dialogism}). Therefore, with the development of DH research technology and interaction with the original humanities and communities. It can be speculated that there should be three stages of ``diversion of technology and theory'', ``dialogue of technology and theory'' and ``fusion of technology and theory''. Because this research object belongs to the literature interpretation in the traditional research field, and it is still a developing DH research environment which this author is located. Therefore, this article expects its own position from ``diversion of technology and theory'' to ``dialogue of technology and theory'', which is also spirit ``CS+X''.

We apply a ``topic model'' methodology, and use Content Analysis Toolkit for Academic Research (CATAR) (see~\cite{tseng2009comparison, tseng2010generic}), which begins with a few basic assumptions:
\begin{enumerate}
\item  Every article (Document A, Document B, etc. in the figure below) is made up of its smallest meaningful units, words. This does not necessarily mean individual Chinese characters.
\item Looking at both which words co-occur across articles and the frequency of such co-occurrence, we can derive a numeric term that describes the level of similarity in material covered between articles.
\item These numeric terms allow us to compare levels of similarity with greater precision and objectivity than simply reading every article personally.
\end{enumerate}

Documents which are more similar, will be clustered together easier. We can pull out the ``co-words'' (words co-occurring in multiple articles) in multiple levels. These ``co-words'' tell us the most important keywords in each cluster of articles. We call these most important co-words ``topics'', and we use them to build ``topic models''. The more samples included in the analysis, the more levels the topic model will contain.

Our preliminary research applied this methodology to all of the articles and essays published in the Taiwanese journal titled Mainland China Studies from 1998 to 2015, which were first collected in a database. Then, We constructed topic models using the titles and summaries of
each piece. This helped identify the most popular research topics and their keywords, in order to more closely examine the trends in each research topic.

\section{Experimental Results}
The results highlight seven main themes in the 473 pieces published by Mainland China Studies between 1998 and 2015, each theme with its own keywords.
Examining the distribution of articles on each topic over time (including both numbers and percentages of articles), we identified two major and enduring topics of the journal (economics/trade and the CCP), in addition to the many other topics that varied over time.

\begin{figure}[!h]
\centerline{\includegraphics[width=1\textwidth]{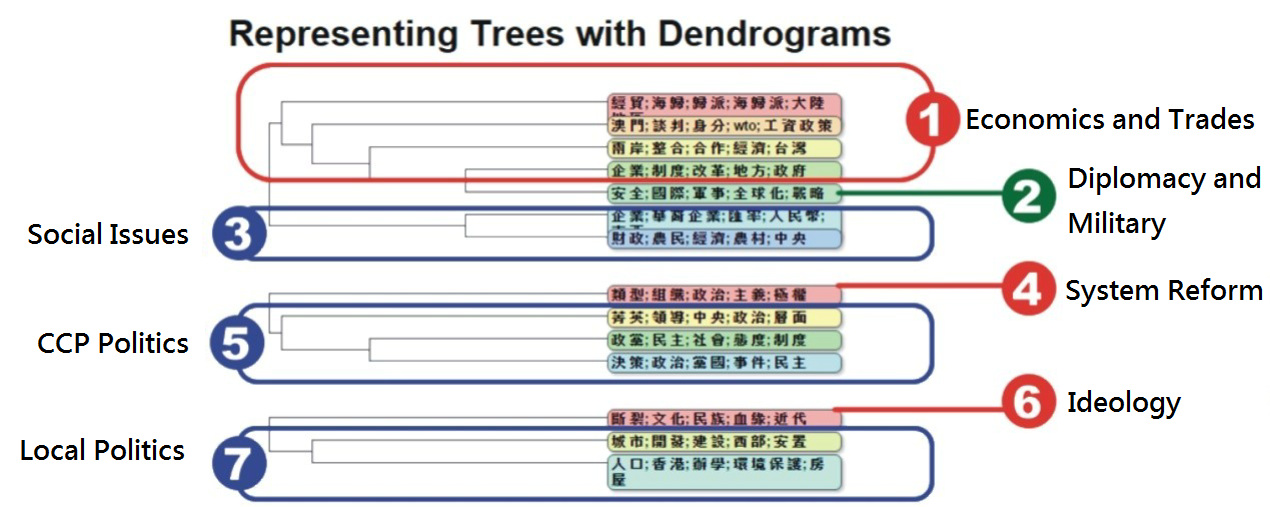}}
\caption{SMD of China Studies in Taiwan}\label{f7}
\end{figure}

Our method outputs identified three major clusters.
The keywords are associated with those clusters.
We might characterize them as clusters around economic or social issues, political systems, and political thought.
But Figure~\ref{f7} shows that we look more closely at the keywords and relying on our domain knowledge in the field of CSIT,
we further divided the three clusters into seven smaller clusters.
These seemed to reflect the distinct themes in each cluster (the first large cluster splitting into three smaller ones, and the two other large clusters each splitting into two smaller ones), as seen below.
Certainly, other researchers might choose to divide the machine's outputted clusters differently or not at all.

\begin{figure}[!h]
\centerline{\includegraphics[width=1\textwidth]{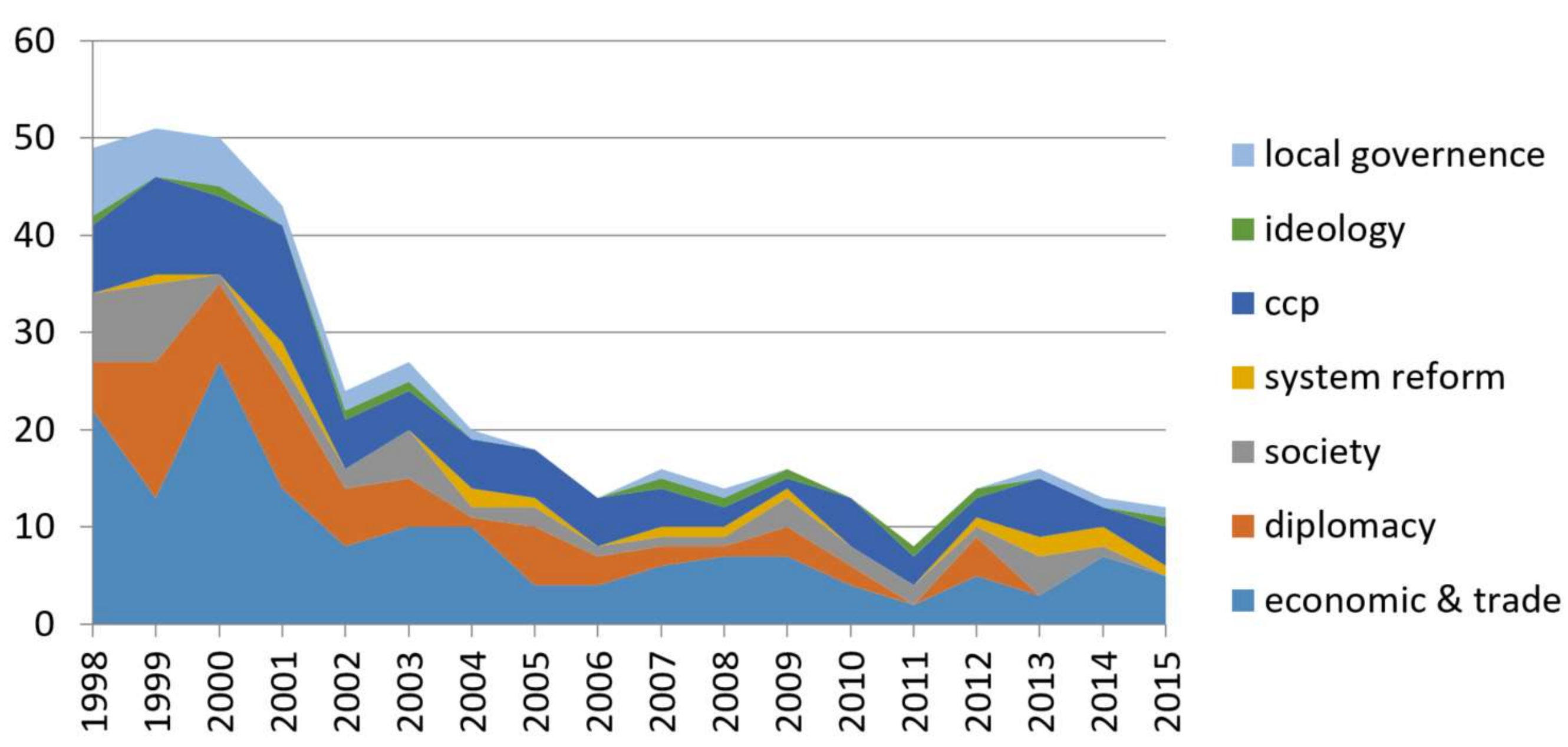}}
\caption{Line Chart of Article Numbers by Year}\label{f8}
\end{figure}

\begin{figure}[!h]
\centerline{\includegraphics[width=1\textwidth]{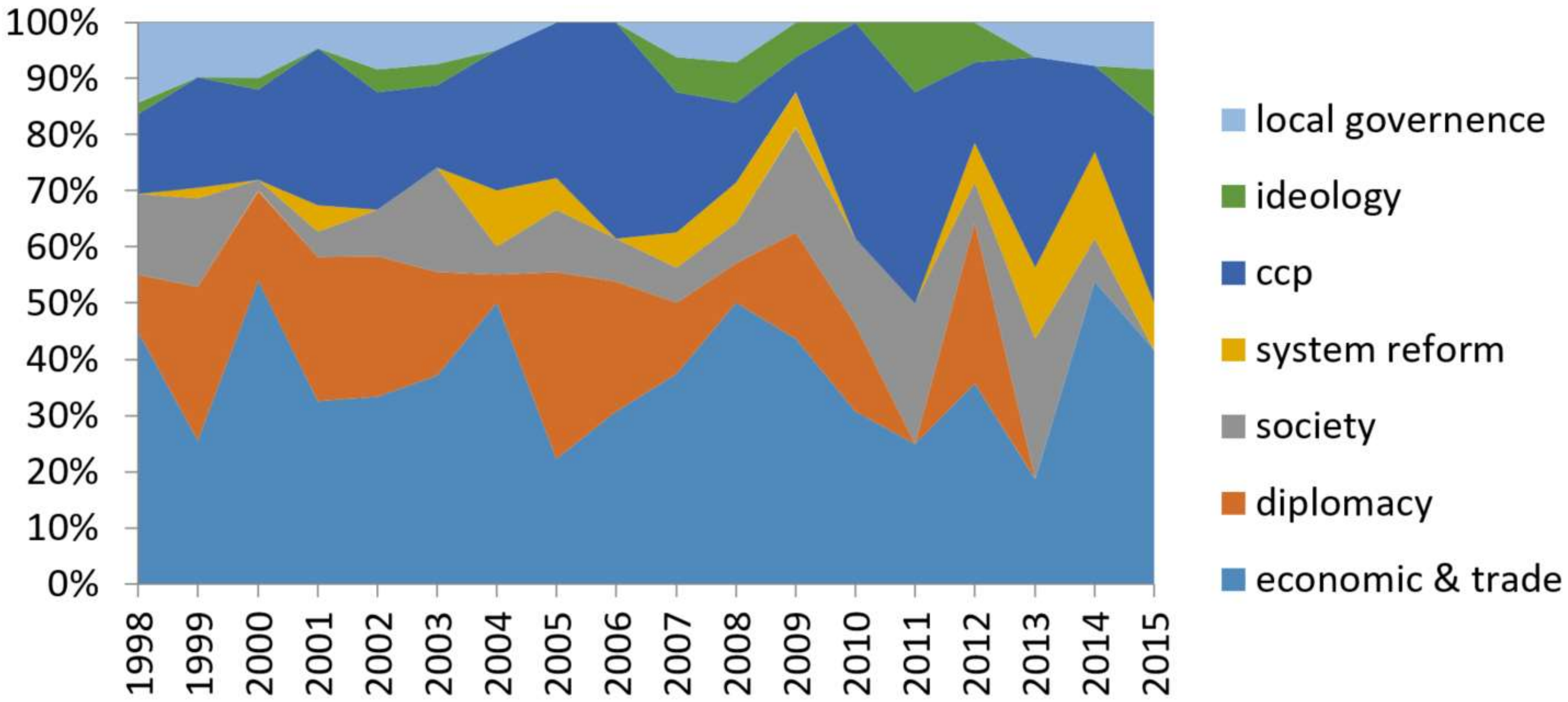}}
\caption{Line Chart of Article Percentages by Year}\label{f9}
\end{figure}

Using that information, Table~\ref{t1} shows a time series of common themes and the number of articles Mainland China Studies published on them from 1998 through 2015. Figure~\ref{f8} shows that the same information can also be visualized by frequency. Figure~\ref{f9} shows that the same information can also be visualized by percentage.

Clearly, the primary theme is ``Economics and Trade'' which has relatively steadily accounted for over a third of all published articles. The keywords for this cluster are ``economy and trade, negotiation, cross-strait, enterprise, system''.
It therefore seems likely that the most frequently researched topic for Taiwanese scholars publishing in Mainland China Studies since 1998 relates to doing business with mainland China. The next biggest cluster is the CCP.
Its keywords are ``politics, leadership, decision-making, movement, party and state''.
This cluster is about political leaders, major political events, and the political party system of the CCP.
We can also see that the cluster for ``Ideology'', never wildly popular, is decreasing in research popularity over time, which is consistent with our experience as a China Studies scholar.
 
These and other clusters can of course be examined in more granular detail by reading the actual articles within them.
We leave that joy of individual discovery to the readers.

\section{Conclusion}
Our research proposes the following contributions for the CSIT, among others:
\begin{enumerate}
\item The text mining tool could help DHer to know their field in bird eye.
\item Our research pointed different stream of CSIT specifically, such as sociology, economics, or other scholarly topics.
\item In the future, the tool can be applied to study non-academic writing, such as political opinions in newspapers, internet comments on Facebook or Twitter, or other sources of information.
\end{enumerate}

With regards of research limitation, the goal of this paper is to demonstrate the effect of using text mining to study Chinese materials on the one hand; on the other hand, it is also hope to increase the interest of the international academic community in the application of Chinese processing techniques.

Therefore, at this stage, more familiar algorithms are used, such as using simple frequency instead of TF-IDF for basic values, and Jaccard distance for similarity, not cosine distance or Euclidean distance.
The comparison of these models can continue to develop in the future.
We also agree that ``all models are wrong, but some are useful'' (see~\cite{grimmer2013text}) and require semantics of the researcher's subject and the use of other methods in the future.

%Contact with Western scholars will be invaluable as we move forward.
%To that end, ERTTC will provide an excellent opportunity for further development, to share our knowledge of digital tools and Taiwanese research methods, to exchange knowledge with researchers from around the world, and to find opportunities for future collaboration.

\section{Tables and Figures}
\subsection{Figure}
%\begin{figure}[!htbp]
%\centerline{\includegraphics[width=1\textwidth]{f1.pdf}}
%\caption{Topic Model Basic Assumptions}\label{f1}
%\end{figure}
%\begin{figure}[!htbp]
%\centerline{\includegraphics[width=1\textwidth]{f2.pdf}}
%\caption{Similarity Map of Documents (SMD)}\label{f2}
%\end{figure}
%\begin{figure}[!htbp]
%\centerline{\includegraphics[width=1\textwidth]{f3.pdf}}
%\caption{Articles, Topics, Clusters}\label{f3}
%\end{figure}
%\begin{figure}[!htbp]
%\centerline{\includegraphics[width=1\textwidth]{f4.pdf}}
%\caption{Database Framework}\label{f4}
%\end{figure}
%\begin{figure}[!htbp]
%\centerline{\includegraphics[width=1\textwidth]{f5.pdf}}
%\caption{Calculation Process}\label{f5}
%\end{figure}
%\begin{figure}[!htbp]
%\centerline{\includegraphics[width=1\textwidth]{f6.pdf}}
%\caption{Outputs}\label{f6}
%\end{figure}

\newpage
\subsection{Table}
\begin{table}[!h]
\small
\centering
\caption{Article Numbers by Year}
\label{t1}
\begin{tabular}{|l|l|l|l|l|l|l|l|l|}
\hline
      & \begin{tabular}[c]{@{}l@{}}Economic \\ \& Trade\end{tabular} & Diplomacy & Society & System Reform & CCP & Ideology & \begin{tabular}[c]{@{}l@{}}Local \\ Governance\end{tabular} & Total \\ \hline
1998  & 22                                                           & 5         & 7       & 0             & 7   & 1        & 7                                                           & 49    \\ \hline
1999  & 13                                                           & 14        & 8       & 1             & 10  & 0        & 5                                                           & 51    \\ \hline
2000  & 27                                                           & 8         & 1       & 0             & 8   & 1        & 5                                                           & 50    \\ \hline
2001  & 14                                                           & 11        & 2       & 2             & 12  & 0        & 2                                                           & 43    \\ \hline
2002  & 8                                                            & 6         & 2       & 0             & 5   & 1        & 2                                                           & 24    \\ \hline
2003  & 10                                                           & 5         & 5       & 0             & 4   & 1        & 2                                                           & 27    \\ \hline
2004  & 10                                                           & 1         & 1       & 2             & 5   & 0        & 1                                                           & 20    \\ \hline
2005  & 4                                                            & 6         & 2       & 1             & 5   & 0        & 0                                                           & 18    \\ \hline
2006  & 4                                                            & 3         & 1       & 0             & 5   & 0        & 0                                                           & 13    \\ \hline
2007  & 6                                                            & 2         & 1       & 1             & 4   & 1        & 1                                                           & 16    \\ \hline
2008  & 7                                                            & 1         & 1       & 1             & 2   & 1        & 1                                                           & 14    \\ \hline
2009  & 7                                                            & 3         & 3       & 1             & 1   & 1        & 0                                                           & 16    \\ \hline
2010  & 4                                                            & 2         & 2       & 0             & 5   & 0        & 0                                                           & 13    \\ \hline
2011  & 2                                                            & 0         & 2       & 0             & 3   & 1        & 0                                                           & 8     \\ \hline
2012  & 5                                                            & 4         & 1       & 1             & 2   & 1        & 0                                                           & 14    \\ \hline
2013  & 3                                                            & 0         & 4       & 2             & 6   & 0        & 1                                                           & 16    \\ \hline
2014  & 7                                                            & 0         & 1       & 2             & 2   & 0        & 1                                                           & 13    \\ \hline
2015  & 5                                                            & 0         & 0       & 1             & 4   & 1        & 1                                                           & 12    \\ \hline
Total & 158                                                          & 71        & 44      & 15            & 90  & 10       & 29                                                          & 417   \\ \hline
\end{tabular}
\end{table}

\newpage
\bibliographystyle{plainnat}
\bibliography{jdmdh-example}
\end{document}